\documentclass{emulateapj}

\usepackage{graphicx,times}
\def\apjl{ApJ}%
\def\apjs{ApJS}%
\def\aap{A\&A}%
\def\mnras{MNRAS}%
\def\gtrsim{\mathrel{\hbox{\rlap{\hbox{\lower4pt\hbox{$\sim$}}}\hbox{$>$}}}}

\shortauthors{H.\ Ebeling, C.-J. Ma, E.\ Barrett}
\shorttitle{Galaxy redshifts in the Frontier Fields}

\begin{document}
\slugcomment{}
\received{September 24, 2013}
\accepted{February 7, 2014}
\title{Spectroscopic redshifts of galaxies within the Frontier Fields}
\author{Harald Ebeling\altaffilmark{1}, Cheng-Jiun Ma\altaffilmark{1,2,3,4}, Elizabeth Barrett\altaffilmark{1}}
\altaffiltext{1}{Institute for Astronomy, University of Hawaii, 2680 Woodlawn Drive, Honolulu, HI 96822, USA}
\altaffiltext{2}{Department of Physics \& Astronomy, University of Waterloo, 200 University Ave.\ W., Waterloo, Ontario, N2L\,3G1, Canada}
\altaffiltext{3}{Harvard-Smithsonian Center for Astrophysics, 60 Garden St., Cambridge, MA, 021380-1516, USA}
\altaffiltext{4}{Institute for Computational Cosmology, Durham University, South Road, Durham, DH1\,3LE, UK}

\begin{abstract}
We present a catalog of 1921 spectroscopic redshifts measured in the fields of the massive galaxy clusters MACSJ0416.1--2403 ($z=0.397$), MACSJ0717.5+3745 ($z=0.546$), and MACSJ1149.5+2223 ($z=0.544$), i.e., three of the four clusters selected by STScI as the targets of the  Frontier Fields (FF) initiative for studies of the distant Universe via gravitational lensing. Compiled in the course of the MACS project (Massive Cluster Survey) that detected the FF clusters, this catalog is provided to the community for three purposes: (1) to allow the identification of cluster members for studies of the galaxy population of these extreme systems, (2) to facilitate the removal of unlensed galaxies and thus reduce shear dilution in weak-lensing analyses, and (3) to improve the calibration of photometric redshifts based on both ground- and spacebased observations of the FF clusters. 
\end{abstract}

\keywords{galaxies:~clusters:~individual (MACSJ0416.1--2403, MACSJ0717.5+3745, MACSJ1149.5 +2223);
galaxies: distances and redshifts
}

\section{\normalsize Introduction}
Recognizing the importance of massive galaxy clusters as gravitational telescopes for studies of the distant universe as originally proposed by the Clusters as Telescopes team (CATs\footnote{Led by Jean-Paul Kneib, the CATs team consists of cluster and lensing experts from the US and Europe; CATs was one of six teams selected by STScI in the spring of 2013 to calibrate the FF cluster lenses.}), Space Telescope Science Institute (STScI) recently announced the Frontier Fields (FF) project, a commitment of 560 orbits of \textit{Hubble Space Telescope (HST)} time in Cycles 21 and 22 to deep observations of four clusters of galaxies at intermediate redshift ($0.3<z<0.6$). Each target will be observed with  both the Advanced Camera for Surveys (ACS) and the Wide Field Camera 3 (WFC3) for 20 orbits in each of the F435W, F606W, F814W, F105W, F125, F140, and F160W filters, reaching a limiting magnitude of approximately 29 (AB system) across the full optical and near-infrared (NIR) window. Both ACS and WFC3 will be used simultaneously, such that for each cluster target images of equal depth are accumulated of the cluster core and of a flanking parallel field about 6 arcmin away. Extending the NIR spectral coverage further,  1000 hours of Spitzer Space Telescope time are being devoted to observations of the same four fields at 3.6 and 4.5$\mu$m. 

The FF project represents the  largest investment of  \textit{HST} time for deep observations of galaxy clusters in the history of  \textit{HST}. Its main science goal is the discovery of gravitationally lensed background galaxies at $z=5-10$ for both in-depth studies of bright individual objects and statistical investigations into the properties of galaxies at magnitudes and distances inaccessible to observation without gravitational amplification. As a bonus, the FF observations will also yield the deepest images of the cores of massive clusters ever collected, thereby facilitating exquisitely detailed characterizations of the cluster lenses and their galaxy content.

The four clusters selected for the FF initiative for Cycle 21 and 22 are A2744 ($z{=}0.308$), MACSJ0416.1--2403 ($z{=}0.397$), MACSJ0717.5+3745 ($z{=}0.546$), and MACSJ1149.5+2223 ($z{=}0.544$) \citep{1958ApJS....3..211A,2003MNRAS.339..913E,2007ApJ...661L..33E,2010MNRAS.407...83E,2012MNRAS.420.2120M}. Subject to a review of the observations collected up to then, two additional clusters (AS1063 and A370) will be added to the FF sample and observed in Cycle 23. All FF clusters are exceptionally X-ray luminous systems ($L_{\rm X}>2\times 10^{45}$ erg s$^{-1}$, bolometric) and feature pronounced substructure, i.e., they combine the two characteristics that select the most powerful cluster lenses \citep{2010MNRAS.406.1318H,2012A&A...544A..71L,2013ApJ...762L..30Z}. Additional information on the FF project can be found on \url{http://www.stsci.edu/hst/campaigns/frontier-fields/}.

\section{\normalsize Characterizing the FF lenses}
If the scientific potential of the FF project for the characterization of the high-redshift Universe is to be fully exploited, the magnification of the cluster lenses and thus their mass distributions must be accurately mapped. Within the cluster cores, the mass distribution can be derived from carefully identified strong-gravitational lensing features, with the greatest leverage being provided by sets of multiple-image systems \citep[and references therein]{2011A&ARv..19...47K}. In addition to cluster-scale halos, the resulting mass models also include appropriately scaled mass halos at the location of the most luminous cluster galaxies, thereby accounting for small-scale perturbations. On larger scales, i.e., outside the strong-lensing regime, complementary constraints on the mass and magnification maps are obtained from measurements of the weak gravitational shear induced by the cluster lens. For the FF clusters, the weak-lensing regime is poorly sampled by the pre-FF, relatively shallow \textit{HST} images; better coverage and thus better constraints on the mass and magnification at larger cluster-centric radii will ultimately emerge from the much deeper FF images themselves (including those of the flanking fields) as well as from supporting wide-field imaging. 

\begin{table*}[t]
\footnotesize 
\begin{tabular}{ccccccp{1.1cm}c}\hline \hline
Run ID &Date & Telescope/Instrument & Grating or &Central & Blocking  & Exposure & Spectral  \\
              &         &                                          & Grism & Wavelength (\AA) & Filter & Time (s) & Resolution (\AA)\\[2mm] \hline
L1 & 2000-11-20/21 & Keck-I/LRIS & 400/3400, 600/7500 & 6200 & GG495 & 3600 (e1) 4800 (e2)& 6.8, 4.7\\
L2 & 2002-11-29 & Keck-I/LRIS & 400/3400, 600/7500 & 6500 & GG495 & 5400 (e1) 6000 (e2) & 6.8, 4.7\\
D1 & 2003-12-23 & Keck-II/DEIMOS & 600ZD & 6500 & GG495 & 7200 & 6.5\\
G1 & 2004-03-12 & Gemini/GMOS & B600 & 6700 & GG455 & 5400 & 4.7\\
G2 & 2004-03-17 & Gemini/GMOS & R400 & 7600 & OG515 & 7200 & 6.9\\
D2 & 2004-12-16  & Keck-II/DEIMOS & 600ZD & 7000 & GG455 & 5400 & 5.7\\
D3 & 2005-02-12  & Keck-II/DEIMOS & 600ZD & 7000 & GG455 & 3240 (e1) 7310 (e2) & 5.7\\
D4 & 2006-01-31   &  Keck-II/DEIMOS & 600ZD & 7000 & GG455 & 6047 (e1) 5400 (e2) 4992 (e3) 5100 (e4) & 5.7\\
D5 & 2006-02-01  & Keck-II/DEIMOS & 600ZD & 7000 & GG455 & 3600 (e1)  5400 (e2) & 5.7\\
D6 & 2006-12-22  & Keck-II/DEIMOS & 600ZD & 7000 & GG455 & 5400 & 4.5\\
D7 & 2008-01-05/06/07  & Keck-II/DEIMOS & 600ZD &  6300 & GG455 & 5400 (e1) 7200 (e2) & 4.5\\ \hline
\end{tabular}
\caption{Observing runs and instrumental setup for our spectroscopic follow-up observations of galaxies in the field of MACSJ0717.5+3745. The effective spectral resolution as listed represents the FWHM of an arc line.\label{tab:m0717-runs}}
\end{table*}

\begin{table*}
\footnotesize 
\begin{tabular}{ccccccp{1.1cm}c}\hline \hline
Run ID &Date & Telescope/Instrument & Grating or &Central & Blocking  & Exposure & Spectral  \\
              &         &                                          & Grism & Wavelength (\AA) & Filter & Time (s) & Resolution (\AA)\\[2mm] \hline
G1 & 2003-05-01 & Gemini/GMOS & B600 & 5900 & OG515 & 3600 & 4.7\\
G2 & 2004-03-16 & Gemini/GMOS & R400 & 7600 & OG515 & 7200 & 6.9\\
D1 & 2005-02-12  & Keck-II/DEIMOS & 600ZD & 7000 & GG455 & 5400 & 5.7\\
D2 & 2006-04-30   &  Keck-II/DEIMOS & 600ZD & 7000 & GG455 & 2700 (e1) 4800 (e2) 5400 (e3) & 5.7\\
D3 & 2009-02-26  & Keck-II/DEIMOS & 600ZD & 6700 & GG455 & 2150 (e1)  3600 (e2) & 5.7\\ \hline
\end{tabular}
\caption{Like Table~\ref{tab:m0717-runs} but for our observations of MACSJ1149.5+2223. \label{tab:m1149-runs}}
\end{table*}

In preparation of the forthcoming FF observations with \textit{HST}, STScI assigned in the summer of 2013 six teams the task of generating maps of the mass distribution of, and gravitational amplification provided by, the FF clusters, based on strong- and weak-lensing analyses of the pre-FF \textit{HST} imaging data (e..g, Richard et al., in preparation). Spectroscopic follow-up observations from the ground were critical in this effort, not only to robustly anchor the lens models via spectroscopic redshifts of multiple-image systems, but also to unambiguously identify cluster members among the galaxies in the FF fields. The latter play an important role as small-scale perturbers of the strong-lensing mass models, and are essential also for the elimination of unlensed galaxies in the weak-lensing analysis.

\begin{figure*}
\includegraphics[width=0.49\textwidth,clip,trim=60mm 38mm 60mm 25mm]{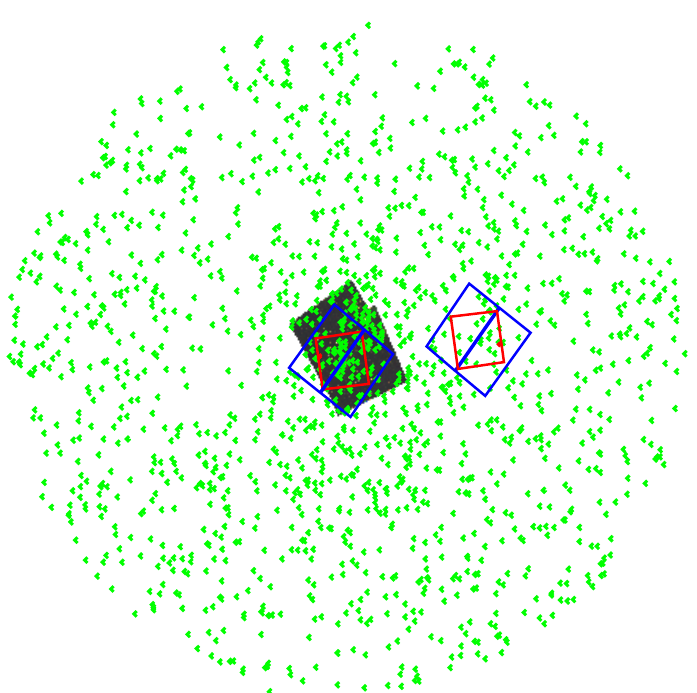}\hfil
\includegraphics[width=0.49\textwidth,clip,trim=35mm 16mm 45mm 5mm]{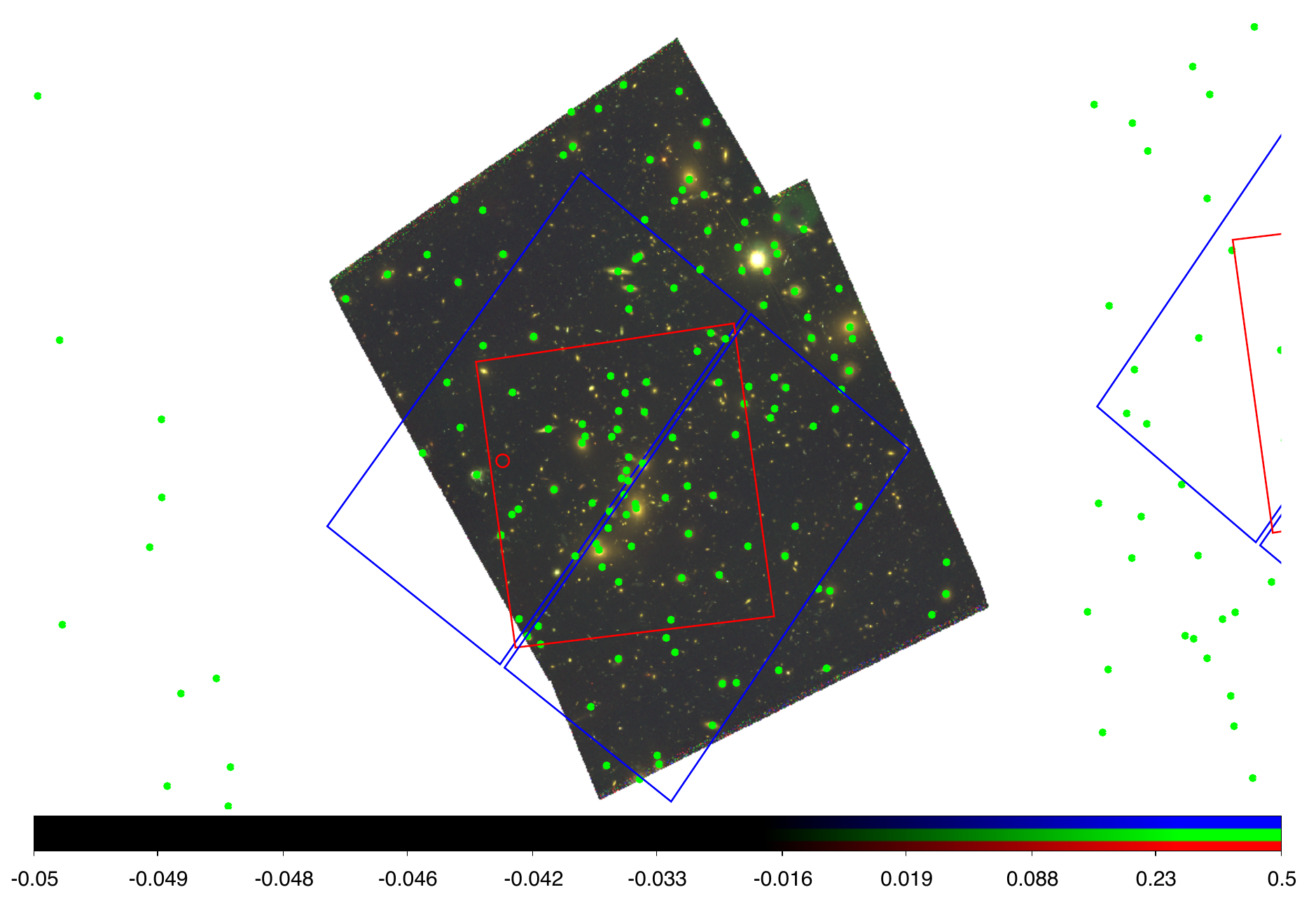}\\
\includegraphics[width=0.49\textwidth,clip,trim=10mm 16mm 60mm 20mm]{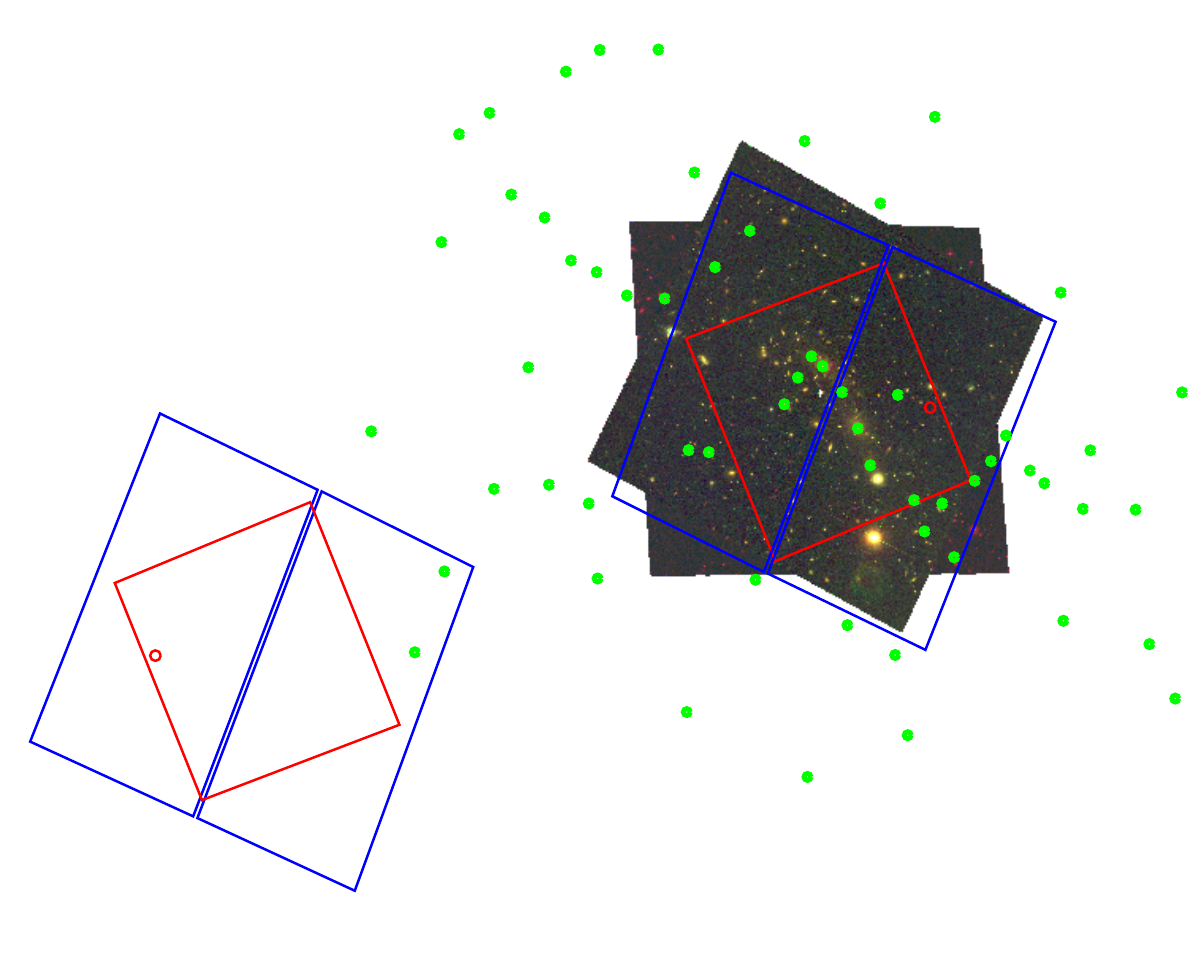}\hfil
\includegraphics[width=0.49\textwidth,clip,trim=35mm 16mm 45mm 5mm]{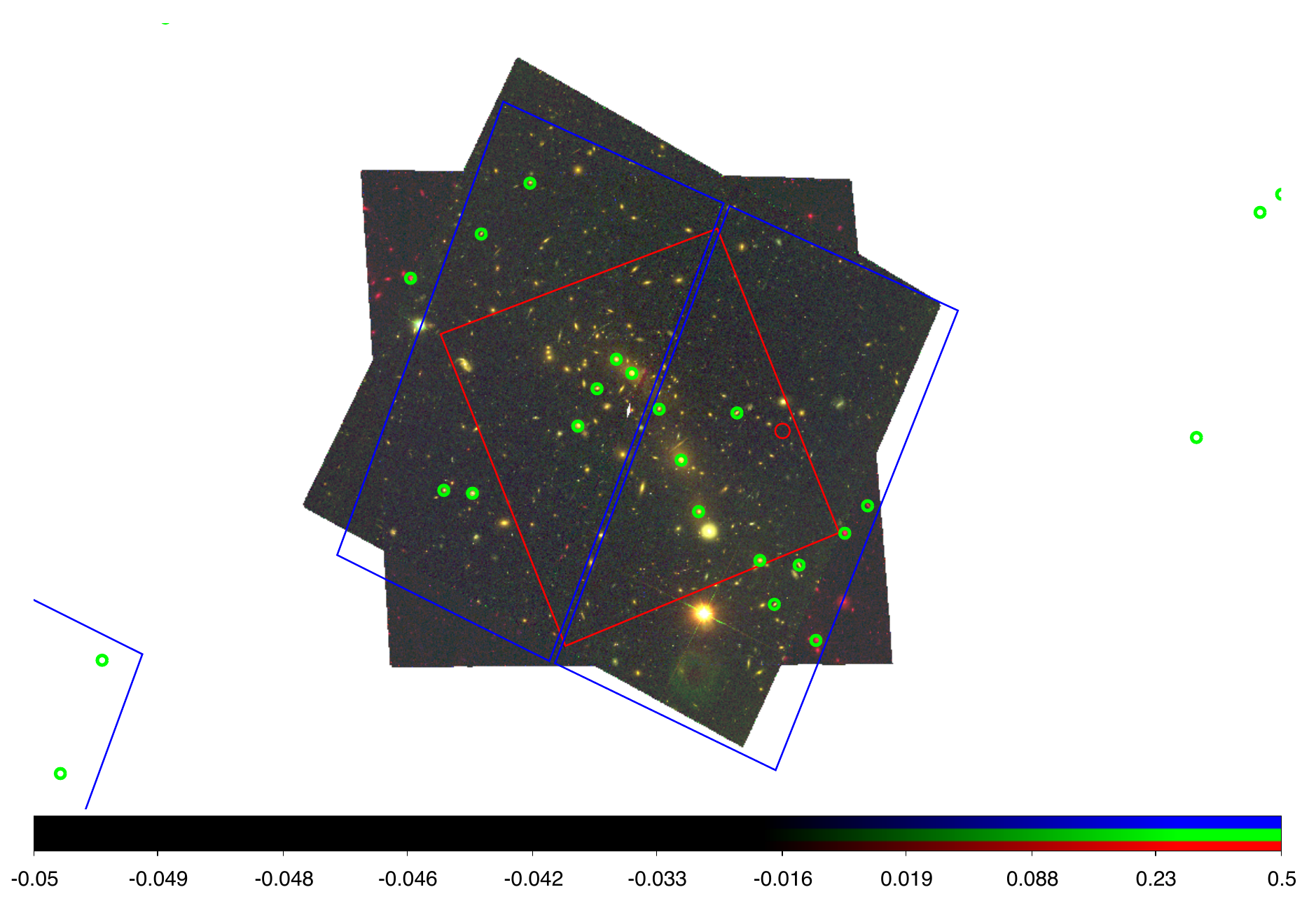}
\caption{Color images (F435W+F606W+F814W) of the archival \textit{HST} data for the FF. Top: A2744, bottom: MACSJ0416.1--2403. Galaxies with spectroscopic redshifts are marked by green circles. For either cluster, the left panel shows the full extent of the spectroscopic data set discussed in Section~\ref{sec:specset}, while the right panel shows a closeup of the data within the area covered by the pre-FF \textit{HST} images of the FF. Overlaid in blue (red) are the apertures of the planned, deep FF observations with ACS (WFC3).\label{fig:FF1}} 
\end{figure*}

\begin{figure*}
\includegraphics[width=0.49\textwidth,clip,trim=55mm 22mm 65mm 35mm]{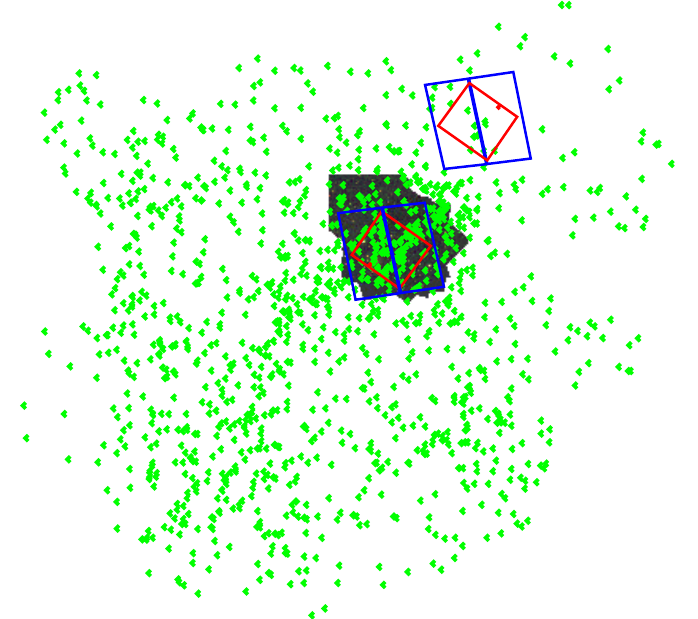}\hfil
\includegraphics[width=0.49\textwidth,clip,trim=35mm 10mm 35mm 0mm]{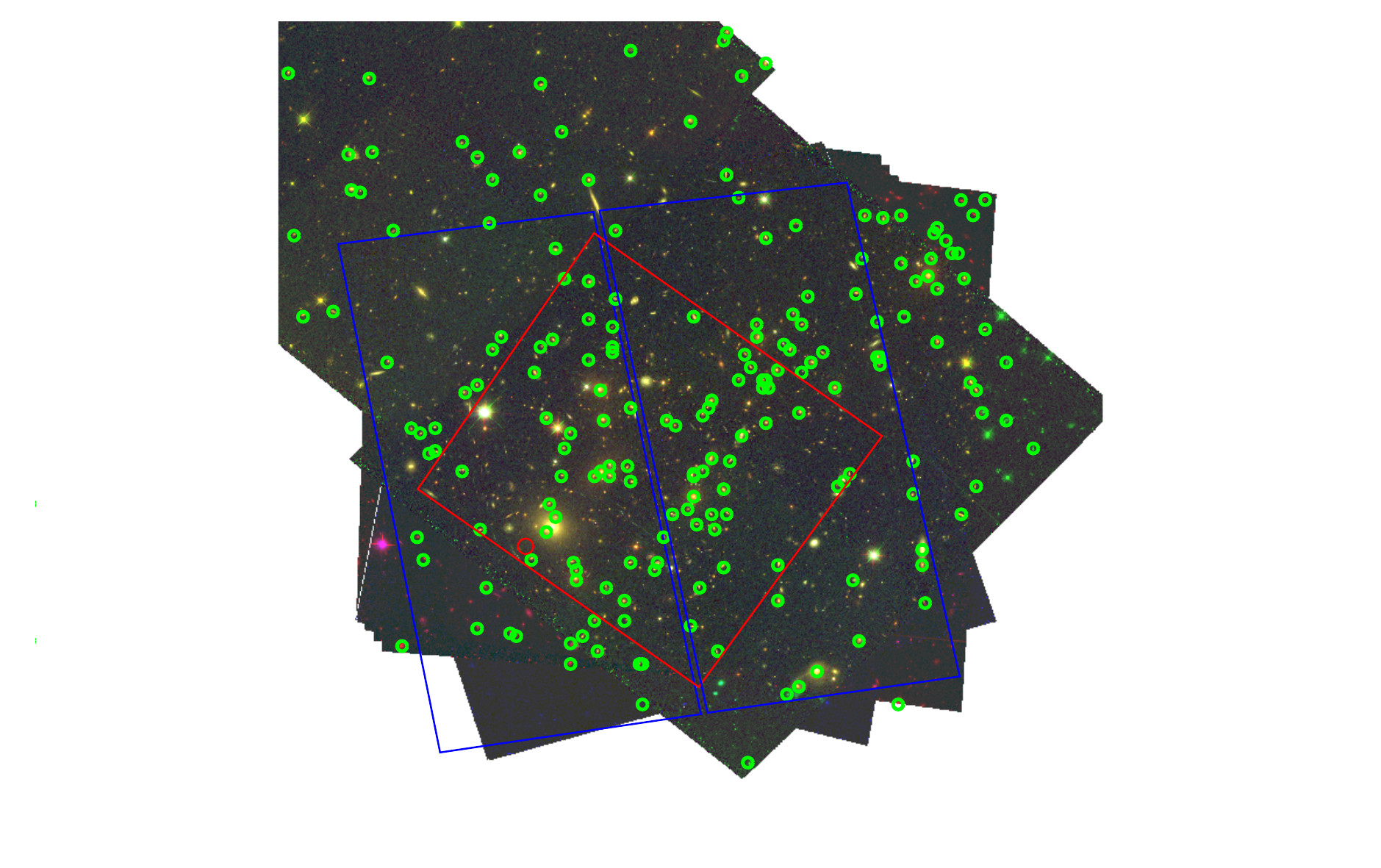}\\
\includegraphics[width=0.49\textwidth,clip,trim=47mm 28mm 75mm 32mm]{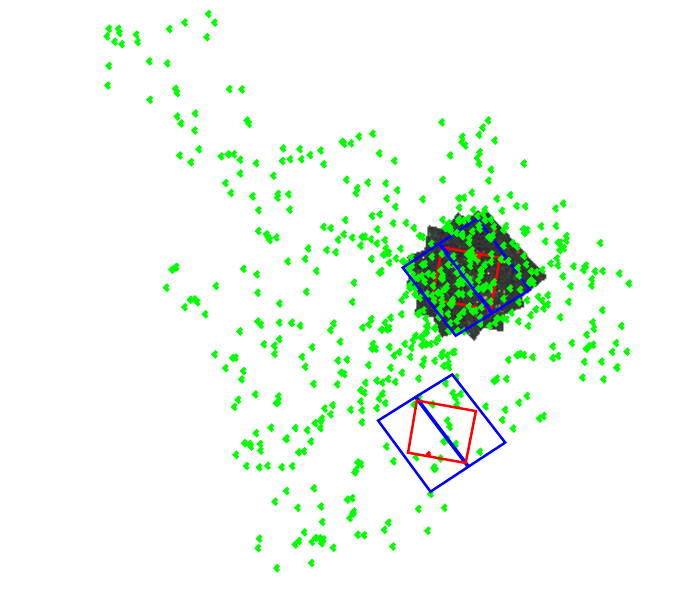}\hfil
\includegraphics[width=0.49\textwidth,clip,trim=30mm 5mm 35mm 3mm]{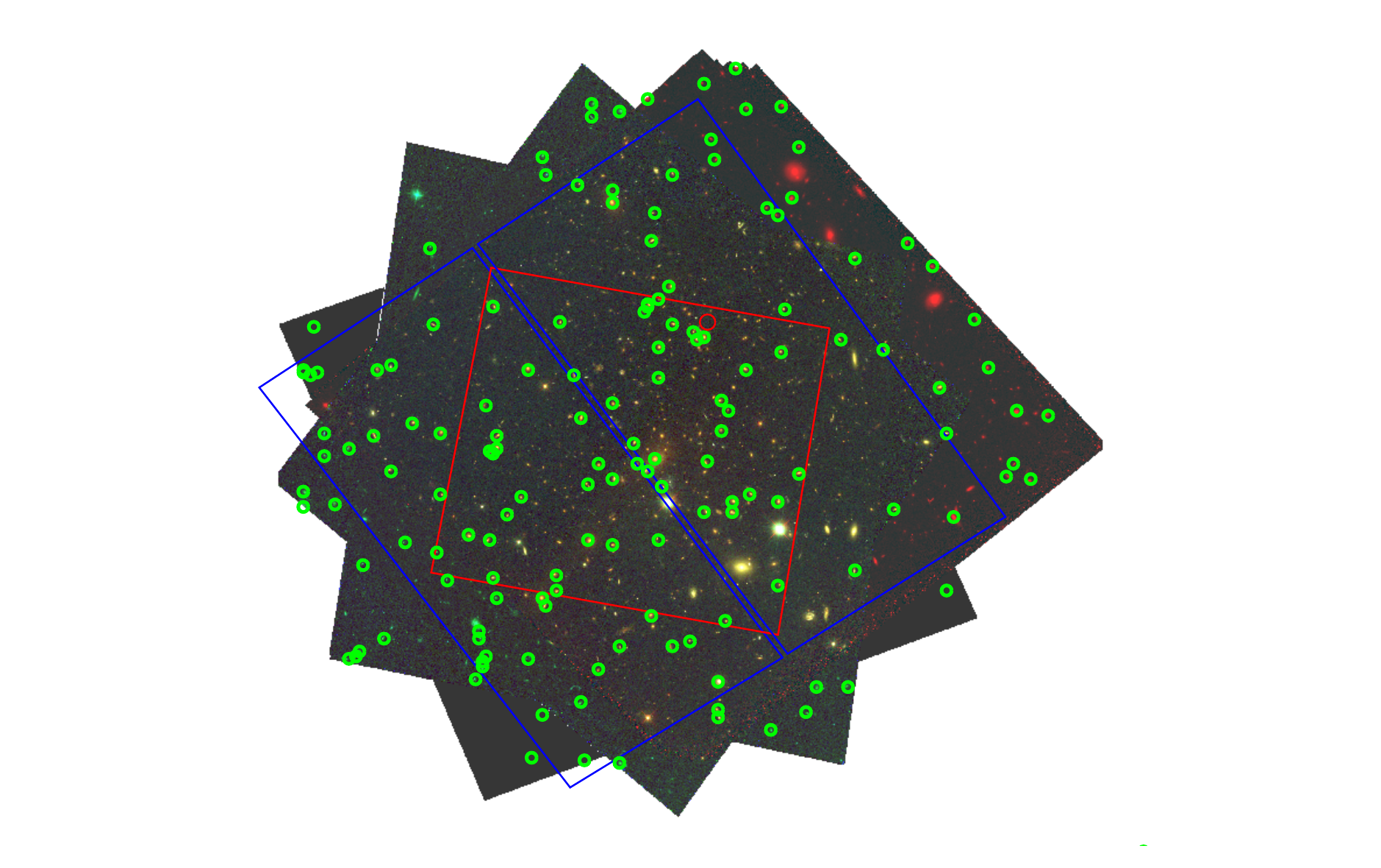}
\caption{As Fig.~\ref{fig:FF1} but for MACSJ0717.5+3745 (top) and MACSJ1149.5+2223 (bottom). The north-eastern quadrant of the image for MACSJ0717.5+3745 includes a small section of the much larger $3\times6$ \textit{HST}/ACS mosaic in the F555W and F814W filters obtained by program GO-10420 (PI Ebeling) and designed to cover the large-scale filament extending to the south-east of the cluster core \citep{2012MNRAS.426.3369J}. \label{fig:FF2}} 
\end{figure*}

In the following we briefly describe spectroscopic follow-up work conducted by us in the fields of three of the four approved FF clusters as part of our extensive spectroscopic survey of the galaxy population of cluster discovered by the Massive Cluster Survey \citep[MACS;][]{2001ApJ...553..668E}. Since the underlying MACS follow-up observations served various scientific purposes, the instrumental settings differed between observing runs,  the spectra obtained vary in resolution and wavelength coverage, and the combined survey results are not statistically complete. The inhomogeneity of this data set notwithstanding, we hope that the information provided here will prove useful for the extragalactic community. 
All redshifts have previously been made available to the six teams generating mass and magnification maps of the FF from pre-FF data. 

\section{\normalsize Spectroscopic observations}\label{sec:specset}
\textit{A2744} \citep[MACSJ0014.3--3022;][]{2010MNRAS.407...83E} was not targeted in the spectroscopic follow-up observations conducted by the MACS team, as the system (also known as AC118) has been covered extensively before \citep{1987MNRAS.229..423C,1998ApJ...497..188C,2006A&A...449..461B,2011ApJ...728...27O}. Galaxies with spectroscopic redshifts  listed in the compilation of \citet{2011ApJ...728...27O} are marked in Fig.~\ref{fig:FF1} (top).

\textit{MACSJ0416.1--2403} was observed by us on January 20, 2001 with the multi-object spectrograph (MOS) on the Canada-Hawaii-France Telescope (CFHT) on Mauna Kea, Hawaii. Our instrumental setup combined the B300 grism with the EEV1 CCD, resulting in a pixel scale of 3.3\AA\  pixel$^{-1}$ and an effective resolution of 12.5\AA\ (as determined by the FWHM of an arc line). We observed a single MOS mask for a total integration time of 6000\,s. Since the goal of these observations was to obtain a credible global velocity dispersion of the cluster, which requires spectra for as many cluster members as possible, the 4504V broadband filter was inserted into the light path, thereby limiting the wavelength range of the resulting spectra to 876\AA\ centred on 5468\AA. The truncated spectra could then be stacked in three to four tiers along the dispersion direction, allowing us to observe many more galaxies on a single MOS mask than would otherwise have been possible. The narrow spectral range covered ensured that redshifts could be obtained from the Ca H+K lines for galaxies at the approximate cluster redshift of $z=0.4$, but did not allow the detection and measurement of most other spectral diagnostics (G band, H$\beta$, [OIII], etc.) 

\textit{MACSJ0717.5+3745} and \textit{MACSJ1149.5+2223} are part of the  subsample of MACS clusters at $z>0.5$ \citep{2007ApJ...661L..33E}, and were observed extensively by us, primarily in the context of a study of the impact of environment on spectral and morphological properties of the cluster galaxy population \citep{2008ApJ...684..160M,2011MNRAS.410.2593M}. We list the respective observing runs and details of the instrumental setup in Tables~\ref{tab:m0717-runs} and \ref{tab:m1149-runs} for our observations of MACSJ0717.5+3745 and MACSJ1149.5+2223, respectively. All observations were performed using multi-object spectrographs on telescopes on Mauna Kea, Hawaii. The vast majority of our spectroscopic data were collected with DEIMOS on Keck-II, using the 600ZD grating. With this setup, the spectral range covered is independent of the chosen central wavelength (which varied between 6300 and 7000\AA); instead, it is determined by the choice of order-blocking filter and by the location of a given slit on the respective MOS mask. For a centrally located slit, the usable wavelength range is approximately 4700 (5200) to 8800\AA\ (9800\AA) for the GG455 (GG495) blocking filter. The effective spectral resolution of our DEIMOS data is typically 6\AA. Spectra were also obtained using LRIS on Keck-I, using the 400/3400 grism in the blue arm, and the 600/7500 grating in the red arm of the spectrograph. Finally, MACSJ0717.5+3745 and MACSJ1149.5+2223 were  observed by us with GMOS on Gemini-N, using the EEV detector array as well as the B600 and R400 gratings, which feature comparable efficiency at 6000 to 7000\AA. Use of the GG495 (OG515) order-blocking filter with the B600 (R400) grating results in a usable spectral range of approximately 5300--8100\AA\ (5400--9600\AA); the spectral coverage is slightly extended or truncated at the blue or red end of this range for slits placed near the edges of the MOS masks. Additional instrumental details, largely similar to the ones used with DEIMOS, can be found in Tables \ref{tab:m0717-runs} and \ref{tab:m1149-runs}.

\section{\normalsize Data reduction}

Generic data reduction procedures were applied to data collected from all telescopes, including bias subtraction, cosmic-ray removal via median averaging of at least three frames where possible, flat fielding, wavelength calibration, skyline straightening, aperture extraction, and background subtraction. 
Following bias subtraction, trimming, and median combining, data for each slit were extracted from the spectral image and analyzed individually thereafter.  Dispersion solutions were obtained using standard calibration lamps (Cu, Cd, Zn, Hg, Kr, Ne, Ar). Skyline straightening was performed when required, and dispersion solutions were verified via comparison to telluric emission lines.  Finally, object spectra were extracted for each aperture and a locally determined background was subtracted. 

For data obtained with CFHT/MOS these reduction steps were performed using the respective \textit{IRAF} tools; for data obtained with Keck and Gemini, use was made of the telescope-specific  Keck-II/DEIMOS pipeline developed by the DEEP2 team \citep{2012ascl.soft03003C,2013ApJS..208....5N} and its adaptation Low-Redux (for Keck-I/LRIS and Gemini/GMOS).

 Since multi-slit masks were used for all observations, flux calibration using standard stars cannot trivially be applied, and redshifts were measured from absorption and emission features using cross correlations of the observed spectra with spectral templates \citep[e.g.,][]{2005A&A...439..845L}. In all cases, the resulting redshifts were verified manually using at least two prominent spectral features, such as (in absorption) Ca H and K, H$\delta$, or the G band, and (in emission) [O II] $\lambda$3727, H$\beta$, and [O III] $\lambda\lambda$4959, 5007. 
 
The accuracy of our redshift measurements is limited by the signal-to-noise ratio (S/N) of our spectra as well as by the precision of their wavelength calibration.  The latter introduces a systematic uncertainty $dz$ of about 0.0002.  We estimate the impact of S/N by comparing the redshifts measured for over 150 galaxies that were observed more than once and feature spectra of comparable S/N. Accounting for the noise in the observed correlation between S/N and redshift differential between repeat measurements, we group all spectra into three broad S/N classes. The standard deviations of the distribution of  redshift differentials for these three groups are measured to be $dz{=}$0.0003, 0.0005, and 0.001, and are consequently assigned as the empirical, 1$\sigma$ redshift uncertainties to spectra with high, medium, and low S/N.

\section{\normalsize Results}

Table~\ref{tab:m0416} lists the coordinates, redshifts, redshift uncertainties, and spectral types (absorption-line, emission-line, and E+A spectra) of the 65 galaxies successfully observed by us with CFHT in the field of MACSJ0416.1--2403. The results from our spectroscopic survey of galaxies in the fields of MACSJ0717.5+3745 (1266 redshifts) and MACSJ1149.5+2223  590 redshifts) are shown in the same format in Tables~\ref{tab:m0717} and \ref{tab:m1149} for the first 20 galaxies (in R.A.\ order) in each field. The full set of results can be obtained electronically from the ApJ website.

\begin{table*}
\tiny
\begin{tabular}{cccc|cccc|cccc}
name & $z$ & DQ & ST & name & $z$ & DQ & ST & name & $z$ & DQ & ST \\[2mm] \hline 
MACSg\,J041549.35-240512.6 & 0.3050 &   3 & A & MACSg\,J041604.82-240538.8 & 0.3844 &   2 & A & MACSg\,J041614.82-240722.7 & 0.3947 &   2 & A\\
MACSg\,J041549.59-240250.2 & 0.3965 &   1 & A & MACSg\,J041605.26-240520.7 & 0.3944 &   1 & A & MACSg\,J041615.74-240324.7 & 0.3121 &   1 & A\\
MACSg\,J041550.24-240257.6 & 0.2817 &   2 & A & MACSg\,J041605.53-240736.0 & 0.3902 &   1 & A & MACSg\,J041616.00-240101.4 & 0.2754 &   2 & A\\
MACSg\,J041550.35-240710.5 & 0.3761 &   2 & A & MACSg\,J041605.95-240420.1 & 0.3976 &   1 & A & MACSg\,J041617.33-240323.0 & 0.4058 &   3 & A\\
MACSg\,J041552.15-240430.1 & 0.4031 &   3 & A & MACSg\,J041606.06-240649.9 & 0.3961 &   1 & A & MACSg\,J041618.47-240101.7 & 0.3896 &   3 & A\\
MACSg\,J041553.00-240717.0 & 0.3560 &   1 & A & MACSg\,J041606.67-240229.9 & 0.3974 &   3 & A & MACSg\,J041618.57-240605.8 & 0.3574 &   1 & A\\
MACSg\,J041553.98-240418.6 & 0.4002 &   3 & A & MACSg\,J041607.09-240500.6 & 0.3970 &   1 & A & MACSg\,J041618.61-240309.5 & 0.3897 &   1 & A\\
MACSg\,J041554.27-240715.0 & 0.3553 &   1 & A & MACSg\,J041607.62-240439.5 & 0.3963 &   1 & A & MACSg\,J041618.93-240522.6 & 0.2806 &   1 & A\\
MACSg\,J041555.36-240643.6 & 0.3937 &   1 & A & MACSg\,J041608.05-240632.6 & 0.3988 &   1 & A & MACSg\,J041619.68-240302.8 & 0.4000 &   1 & A\\
MACSg\,J041555.94-240526.3 & 0.3965 &   1 & A & MACSg\,J041608.28-240418.5 & 0.4008 &   2 & A & MACSg\,J041619.90-240114.1 & 0.4004 &   3 & A\\
MACSg\,J041557.83-240452.0 & 0.4002 &   3 & A & MACSg\,J041609.10-240403.8 & 0.4002 &   1 & A & MACSg\,J041620.61-240511.8 & 0.3991 &   2 & A\\
MACSg\,J041558.16-240525.8 & 0.3761 &   3 & A & MACSg\,J041609.57-240358.0 & 0.3950 &   1 & A & MACSg\,J041620.79-240238.0 & 0.3918 &   2 & A\\
MACSg\,J041558.98-240630.1 & 0.4089 &   3 & A & MACSg\,J041609.75-240800.1 & 0.3430 &   1 & A & MACSg\,J041621.48-240404.3 & 0.3984 &   1 & A\\
MACSg\,J041559.10-240321.3 & 0.3904 &   1 & A & MACSg\,J041609.85-240154.1 & 0.4004 &   3 & A & MACSg\,J041622.19-240224.8 & 0.3890 &   1 & A\\
MACSg\,J041559.78-240511.0 & 0.3070 &   1 & A & MACSg\,J041610.15-240410.3 & 0.3996 &   1 & A & MACSg\,J041622.92-240514.3 & 0.3972 &   1 & A\\
MACSg\,J041600.38-240503.7 & 0.3072 &   1 & A & MACSg\,J041610.72-240425.5 & 0.4031 &   3 & A & MACSg\,J041623.10-240137.7 & 0.3898 &   3 & E\\
MACSg\,J041601.39-240443.6 & 0.3945 &   1 & A & MACSg\,J041611.92-240606.5 & 0.3926 &   1 & A & MACSg\,J041624.38-240150.1 & 0.4012 &   3 & A\\
MACSg\,J041602.02-240458.3 & 0.4016 &   2 & A & MACSg\,J041612.16-240245.8 & 0.3865 &   1 & A & MACSg\,J041625.01-240601.7 & 0.3960 &   1 & A\\
MACSg\,J041602.70-240509.6 & 0.3962 &   1 & A & MACSg\,J041613.62-240306.7 & 0.3923 &   1 & A & MACSg\,J041625.13-240252.3 & 0.3981 &   1 & A\\
MACSg\,J041603.57-240553.5 & 0.4000 &   1 & A & MACSg\,J041613.89-240453.2 & 0.3533 &   1 & E & MACSg\,J041626.26-240648.3 & 0.4123 &   2 & A\\
MACSg\,J041604.07-240522.7 & 0.3960 &   1 & A & MACSg\,J041614.50-240212.1 & 0.4020 &   3 & A & MACSg\,J041628.08-240441.1 & 0.4003 &   1 & A\\
MACSg\,J041604.39-240140.2 & 0.3967 &   1 & A & MACSg\,J041614.74-240451.9 & 0.3118 &   1 & A & \\
\end{tabular}
\caption{Spectroscopic redshifts measured at CFHT for 65 galaxies in the field of MACSJ0416.1--2403. The final 18 characters of the name represent the object's equatorial coordinates for the J2000 equinox. 
The data quality (DQ) indicator in the third column provides an estimate of the measurement accuracy; a value of 1, (2, 3) translates into a 1$\sigma$ uncertainty of approximately 0.0003 (0.0005, 0.001) in the redshift of the respective galaxy. The spectral type (ST) is listed in the final column; we distinguish between absorption-line spectra (A), emission-line spectra (E), and E+A galaxies (E+A).\label{tab:m0416}}
\end{table*}

\begin{table}
\begin{tabular}{ccccc}
name & $z$ & DQ & ST & Instrument \\[2mm] \hline 
MACSg J071639.04+374808.3&  0.3498&   1&    E&    D\\
MACSg J071642.04+374852.6&  0.3349&   1&    E&    D\\
MACSg J071644.13+374601.5&  0.3349&   1&    A&    D\\
MACSg J071644.50+374921.0&  0.3351&   1&    E&    D\\
MACSg J071644.58+374542.2&  0.2603&   1&    A&    D\\
MACSg J071644.81+374734.5&  0.4610&   1&    A&    D\\
MACSg J071644.93+374900.8&  0.3358&   1&  E+A&    D\\
MACSg J071645.64+374123.5&  0.5426&   1&    A&    D\\
MACSg J071645.96+374622.3&  0.6130&   1&    E&    D\\
MACSg J071647.05+374008.1&  0.5382&   1&    E&    D\\
MACSg J071647.34+374107.7&  0.5436&   1&    E&    D\\
MACSg J071647.59+374007.7&  0.3665&   1&    E&    D\\
MACSg J071647.91+374652.6&  0.4611&   1&    E&    D\\
MACSg J071648.22+374540.5&  0.4023&   1&    E&    D\\
MACSg J071649.25+375121.7&  0.5137&   2&    E&    D\\
MACSg J071649.35+374018.2&  0.5433&   1&    A&    D\\
MACSg J071649.45+374544.4&  0.5150&   1&    E&    D\\
MACSg J071650.86+374618.5&  0.3352&   2&    E&    D\\
MACSg J071651.01+374206.9&  0.3343&   1&    E&    D\\
MACSg J071651.40+375235.1&  0.4200&   2&    E&    D\\
\end{tabular}
\caption{Spectroscopic redshifts for 1266 galaxies in the field of MACSJ0717.5+3745. Columns as in Table~\ref{tab:m0416}, except that we also list the instrument used (D for Keck/DEIMOS, L for Keck/LRIS, G for Gemini-N/GMOS). We list only the first 20 galaxies; the full set is available online. \label{tab:m0717}}
\end{table}

\begin{table}
\begin{tabular}{ccccc}
name & $z$ & DQ & ST & Instrument \\[2mm] \hline 
MACSg J114909.03+222346.2&  0.5961&   3&    E&    D\\
MACSg J114909.38+222107.5&  0.9309&   3&    E&    D\\
MACSg J114910.30+222207.7&  0.5644&   2&    E&    D\\
MACSg J114910.99+222031.6&  0.5654&   1&    E&    D\\
MACSg J114911.06+222030.7&  0.5643&   1&    E&    D\\
MACSg J114911.20+222128.1&  0.9351&   3&    E&    D\\
MACSg J114913.28+222004.3&  0.5673&   2&    E&    D\\
MACSg J114913.40+222414.9&  0.5342&   1&    A&    D\\
MACSg J114913.62+222045.3&  0.4549&   1&    E&    D\\
MACSg J114913.83+222520.3&  0.5495&   1&    A&    D\\
MACSg J114914.74+222413.9&  0.5351&   1&    E&    D\\
MACSg J114914.86+222200.8&  0.6912&   2&    E&    D\\
MACSg J114914.97+222123.7&  0.4585&   1&    A&    D\\
MACSg J114915.11+222215.9&  0.5471&   2&  E+A&    D\\
MACSg J114915.19+222355.8&  0.5339&   1&    E&    D\\
MACSg J114915.33+222341.0&  0.5340&   2&    E&    D\\
MACSg J114915.71+222120.7&  0.4558&   3&    A&    D\\
MACSg J114916.15+222147.3&  0.6639&   2&    A&    D\\
MACSg J114916.19+222114.5&  0.5422&   1&    A&    D\\
MACSg J114916.38+222425.7&  0.5429&   2&    A&    D\\
\end{tabular}
\caption{Spectroscopic redshifts for 590 galaxies in the field of MACSJ1149.5+2223. Columns as in Table~\ref{tab:m0717}. We list only the first 20 galaxies; the full set is available online. \label{tab:m1149}}
\end{table}

\section*{Acknowledgements}

HE gratefully acknowledges financial support from STScI grants GO-9722 and GO-10420. The \textit{HST} images of the FF shown in Figs.~\ref{fig:FF1} and \ref{fig:FF2} were reduced and kindly made available by Anton Koekemoer.

\bibliographystyle{plain}

\begin{thebibliography}{}
\bibitem[\protect\citeauthoryear{Abell}{1958}]{1958ApJS....3..211A} G.~O.\ Abell, 1958, {\em \apjs}, 3, 211
\bibitem[\protect\citeauthoryear{Boschin et al.}{2006}]{2006A&A...449..461B} Boschin W., Girardi M., Spolaor M., Barrena R., 2006, A\&A, 449, 461 
\bibitem[\protect\citeauthoryear{Cooper et al.}{2012}]{2012ascl.soft03003C} Cooper M., Newman J.A., Davis M., Finkbeiner D.P., Gerke B.F., 2012, ASCL, 1203.003
\bibitem[\protect\citeauthoryear{Couch \& Sharples}{1987}]{1987MNRAS.229..423C} Couch W.J.\ \& Sharples R.M., 1987, MNRAS, 229, 423
\bibitem[\protect\citeauthoryear{Couch et al.}{1998}]{1998ApJ...497..188C} Couch W.J., Barger A.J., Smail I., Ellis R.S., Sharples R.M., 1998, ApJ, 497, 188
\bibitem[\protect\citeauthoryear{Ebeling et al.}{2001}]{2001ApJ...553..668E} Ebeling H., Edge A.C., Henry J.P., 2001, ApJ, 553, 668
\bibitem[\protect\citeauthoryear[{Ebeling et al.}{2007}]{2007ApJ...661L..33E} Ebeling H., Barrett E.,  Donovan D., Ma C.-J., Edge A. C., Van Speybroeck L., 2007, ApJ, 661, L33
\bibitem[\protect\citeauthoryear{Ebeling et al.}{2010}]{2010MNRAS.407...83E} Ebeling H., Edge A.~C., Mantz A., Barrett E., Henry J.~P., Ma C.~J., van Speybroeck L., 2010, MNRAS, 407, 83 
\bibitem[Edge et al.(2003)]{2003MNRAS.339..913E} Edge, A.~C., Ebeling, H., 
Bremer, M., et al.\ 2003, \mnras, 339, 913 
\bibitem[Horesh et al.(2010)]{2010MNRAS.406.1318H} Horesh, A., Maoz, D.,  Ebeling, H., Seidel, G., \& Bartelmann, M.\ 2010, \mnras, 406, 1318 
\bibitem[\protect\citeauthoryear{Jauzac et al.}{2012}]{2012MNRAS.426.3369J}  Jauzac M., et al., 2012, MNRAS, 426, 3369 
\bibitem[\protect\citeauthoryear[{Kneib \& Natarajan}{2011}]{2011A&ARv..19...47K} Kneib J.-P.\ and Natarajan P., 2011, AAPR, 18, 47
\bibitem[Le F{\`e}vre et  al.(2005)]{2005A&A...439..845L} Le F{\`e}vre, O., Vettolani, G., Garilli, B., et al.\ 2005, \aap, 439, 845 
\bibitem[Limousin et  al.(2012)]{2012A&A...544A..71L} Limousin, M., Ebeling, H., Richard, J., et al.\ 2012, \aap, 544, A71 
\bibitem[\protect\citeauthoryear{Ma \& Ebeling}{2011}]{2011MNRAS.410.2593M} Ma C.-J., Ebeling H., 2011, MNRAS, 410, 2593 
\bibitem[\protect\citeauthoryear{Ma et al.}{2008}]{2008ApJ...684..160M} Ma C.-J., Ebeling H., Donovan D., Barrett E., 2008, ApJ, 684, 160 
\bibitem[\protect\citeauthoryear{Mann \& Ebeling}{2012}]{2012MNRAS.420.2120M} Mann A.W.\ \& Ebeling H., 2012, MNRAS, 420, 2120
\bibitem[\protect\citeauthoryear{Newman et al.}{2013}]{2013ApJS..208....5N} Newman J.A.\ et al., 2013, ApJS, 208, 5N
\bibitem[\protect\citeauthoryear{Owers et al.}{2011}]{2011ApJ...728...27O} Owers M.S., Randall S.W., Nulsen P.E.J., Couch W.J., David L.P., Kempner J.C., 2011, ApJ, 728, 27
\bibitem[Zitrin et al.(2013)]{2013ApJ...762L..30Z} Zitrin, A., Meneghetti, M., Umetsu, K., et al.\ 2013, \apjl, 762, L30 

\end{thebibliography}

\end{document}